\documentclass[conference]{IEEEtran}
\IEEEoverridecommandlockouts

\usepackage[noadjust]{cite}
\usepackage{color}
\usepackage{url}
\usepackage{hyperref}
\usepackage[latin9]{inputenc}
\usepackage{epstopdf}
\usepackage{amsthm}
\usepackage{amsmath}
\usepackage{amsthm}
\usepackage{amssymb}
\usepackage{graphicx}
\usepackage{subfigure}
\usepackage{wrapfig}
\usepackage{algorithmic}
\usepackage{algorithm}
\usepackage{mathrsfs}
\usepackage{float}
\usepackage{mathtools}
\usepackage{tabulary}
\usepackage{booktabs}
\usepackage{caption}
\usepackage{multirow}
\usepackage{array}
\usepackage{makecell}
\usepackage{enumitem}

\usepackage[export]{adjustbox}
\usepackage{tabularray}
\PassOptionsToPackage{bookmarks=false}{hyperref}

\usepackage[dvipsnames]{xcolor}

\def\BibTeX{{\rm B\kern-.05em{\sc i\kern-.025em b}\kern-.08em
   T\kern-.1667em\lower.7ex\hbox{E}\kern-.125emX}}

\newcommand{\comment}[1]{ }

\newcommand\subparagraph{%
  \@startsection{subparagraph}{0}
  {\parindent}
  {0ex \@plus 0ex \@minus 0ex}
  {-1em}
  {\normalfont\normalsize\bfseries}}
\makeatother

\usepackage{titlesec}
\titlespacing*{\section}      {0pt}{*0.4}{*0.1}  
\titlespacing*{\subsection}   {0pt}{*0.4}{*0.1}   
\titlespacing*{\subsubsection}{0.9pt}{*0.35}{*0.1}   

\begin{document}

\title{Replicating the Signature: Unsupervised Targeted Impersonation Attack on RF Fingerprinting}

\author{Haytham Albousayri and Bechir Hamdaoui\\
\small School of Electrical Engineering and Computer Science, Oregon State University, Corvallis, OR, USA\\
Emails: \{albousah, hamdaoui\}@oregonstate.edu 
\thanks{This work is supported in part by NSF Award No. 2350214.}}

\maketitle

\thispagestyle{empty}
\pagestyle{empty}

\begin{abstract}
This paper presents a novel impersonation attack framework that aims to fool RF Fingerprinting (RFFP) identification systems by synthesizing signals that replicate the hardware-specific impairments of a target device. Our framework leverages unsupervised learning to enable accurate impairment estimation, combined with signal processing-based generation to synthesize high-fidelity adversarial signals. Unlike prior works that assume full access to the legitimate (victim) RFFP classifier, we consider a more realistic attack strategy where the adversary performs the attack from a completely different transceiver hardware. We further evaluate our proposed attack under realistic and challenging deployment settings, including over-the-air transmission in both Line-of-Sight (LoS) and Non-Line-of-Sight (NLoS) scenarios. Extensive experiments conducted on a Bluetooth Low Energy (BLE) device testbed demonstrate that our attacks remain highly effective even under severe access constraints, significantly outperforming existing baselines in terms of targeted attack success rates by over $80\%$. We additionally analyze the effects of cross-domain generalization, signal representation mismatch, and classifier diversity, highlighting the robustness and transferability of the proposed attack framework.
\end{abstract}



\section{Introduction}
\label{sec:introduction}
Deep learning (DL)-based RF Fingerprinting (RFFP) has become a powerful approach to enhancing physical-layer security. By leveraging device-specific hardware impairments embedded in the received signals---such as carrier frequency offset, I/Q imbalance, and power amplifier nonlinearities, RFFP enables reliable device authentication and identification~\cite{xing2022design}. 
While recent works have substantially improved the scalability and domain generalizability of these RFFP systems~\cite{albousayri2025bluetooth, johnson2025domain,tang2024causal, jagannath2023embedding, xing2022design}, their robustness against active attackers remain largely underexplored. 
In this context, an attacker can compromise the system during deployment in two ways: (i) by transmitting a malicious jamming signal that confuses the classifier and causes it to misidentify the legitimate user, or, more critically, (ii) by crafting and transmitting an impersonation signal that mimics a specific device and is falsely recognized as having originated from that device.
Understanding and addressing the latter threat is essential for ensuring the reliability and security of DL-based RFFP systems in real-world deployment.

This work introduces a novel framework for generating attack signals that replicate impairments of targeted (legitimate) devices, causing a victim RFFP system to misclassify these forged transmissions as authentic.
Unlike prior works that rely on unrealistic assumptions---such as modifying signals at the victim receiver~\cite{liu2023robust}---or are limited to replaying altered versions of captured signals, thereby transmitting the same underlying payload~\cite{danev2010attacks, rehman2014analysis}, our approach generates signals with controllable adversarial payload content while preserving the target device's fingerprints, all without requiring any knowledge of the victim RFFP system.
Extensive experimental results demonstrate that current DL-based RFFP systems remain vulnerable to such impersonation attacks, highlighting the need for more robust and resilient defense mechanisms.


\subsection{Related Work}
Recent works have focused on improving the robustness and generalization of RFFP systems under challenging and unseen conditions~\cite{albousayri2025bluetooth, johnson2025domain,tang2024causal, jagannath2023embedding, xing2022design}. However, their vulnerability to adversarial attacks remains largely unexplored.
Recent studies have begun investigating these vulnerabilities from two primary attacker objectives: \textit{Misclassification}~\cite{yao2026physical, ma2025adversarial, lu2025generative, liu2023robust}, where the adversary injects perturbations to disrupt legitimate devices authentication, and \textit{Impersonation}~\cite{rehman2014analysis, deng2025intrusion, xu2025collusion, merchant2019securing}, where the attacker aims to replicate a specific device's hardware-induced fingerprint such that its transmissions are classified as that target device.


\subsubsection{Misclassification Attacks}
Several works studied the RFFP vulnerability to malicious perturbations inspired by adversarial machine learning~\cite{kurakin2018adversarial}.
For instance, Liu et al.~\cite{liu2023robust} considered adding signal perturbations to induce RFFP misclassification. Lu et al.~\cite{lu2025generative} extended the idea and proposed predicting the optimal adversarial perturbations through DL models. Both studies, however, are input-dependent and assume a prior knowledge of the exact sequence used by the victim classifier.
Subsequently, Ma et al.~\cite{ma2025adversarial} introduced input-independent "universal adversarial perturbations" and demonstrated high misclassification rates.
Despite their effectiveness, all these works primarily report attack success rate without considering its impact on bit error rate (BER). More recently, Yao et al.~\cite{yao2026physical} studied the trade-off between perturbation power and BER and proposed a training-based approach that attacks RFFP classifiers while ensuring communication reliability. While promising, their evaluation was done through simulation, leaving real-world practicality under hardware and channel impairments largely unexplored.

\subsubsection{Impersonation Attacks}
Early studies on RFFP impersonation have primarily focused on replay-based attacks~\cite{rehman2014analysis}, in which adversaries capture legitimate signals and retransmit them (unmodified) using their own transceivers to mimic the target device.
These works showed that the replaying transceiver inevitably imprints its own hardware impairments onto the signal, distorting the original fingerprint and limiting impersonation accuracy. 
Merchant et al.~\cite{merchant2019securing} were among the first researchers who investigated the use of Generative Adversarial Networks (GANs) to synthesize IQ signals, emulating the RF fingerprint of target devices. 
Deng et al.~\cite{deng2025intrusion} improved physical realism by incorporating power amplifier nonlinearity models into the GAN-based generator. However, both works assumed full access to the victim classifier.
Lately, Xu et al.~\cite{xu2025collusion} proposed a collusion-based impersonation framework using a VAE-based generator, where a third party, referred to as a colluder, serves as the discriminator and provides feedback to help the attacker train a raw IQ signal generator. Their approach, however, relied on full knowledge of the victim classifier and was only validated exclusively on simulated MATLAB environments.



\begin{figure}
    \centering
    
    \begin{minipage}{0.55\linewidth} 
    \vspace{-10pt}
        \centering
        \includegraphics[width=\linewidth]{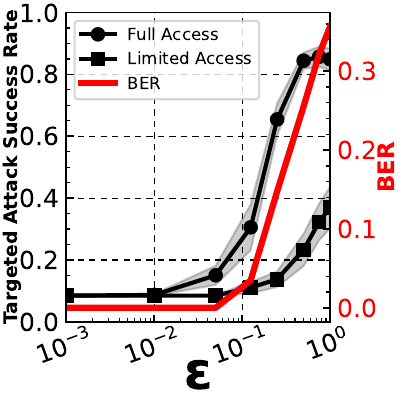}
        
        \vspace{-6pt}
        \SetCell[c=1]{c} {\small \text{(a) TASR vs $\bf{\epsilon}$.}}
        \label{fig:Mot1_v2}
    \end{minipage}
    \hspace{-2.5mm}
    \begin{minipage}{0.405\linewidth}
        \centering
        \includegraphics[width=\linewidth]{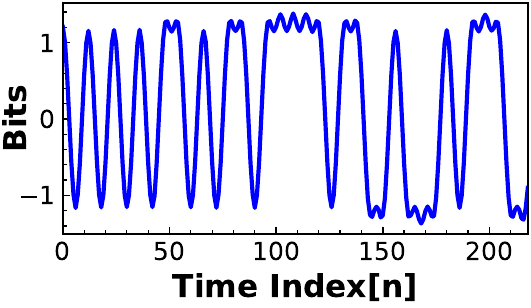}
        
        \SetCell[c=1]{c} {\small \text{(b) Ideal (BER=0\%)}}
        \vspace{+4pt} 
        
        \label{fig:dd}        
        \includegraphics[width=\linewidth]{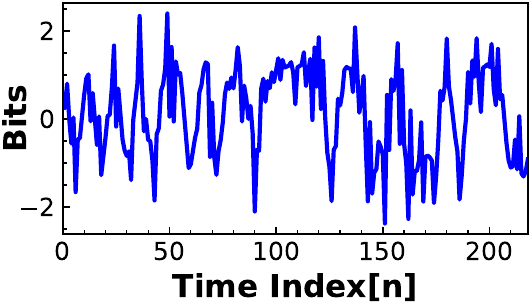}
        
        \SetCell[c=1]{c}{\small \text{(c) $\epsilon=0.125$ (BER=3.5\%)}}
        \label{fig:gg}
    \end{minipage}
    \vspace{-8pt}
     \caption{RFFP baseline attacks on a classifier trained on 12 BLE devices under Full-Access and Limited-Access models.}
    \label{fig:Motivation}
\end{figure}

\subsection{Limitations of Prior Attack Models: Our Motivation}\label{subsec:limitation}
In general, prior work has been evaluated largely under optimistic assumptions, with three key limitations:
\begin{enumerate}[noitemsep, topsep=0pt, leftmargin=*]
    \item {\bf Communication reliability compromise:}
    Prior approaches often overlook the tradeoff between adversarial capability and communication reliability: an impersonation signal must also remain decodable as the targeted device.
    As illustrated in Figure~\ref{fig:Motivation}(a), the widely adopted additive perturbation impersonation attack (bounded by $\epsilon$)~\cite{kurakin2018adversarial} may increase the success rate of targeted attacks, but only at the expense of degraded communication quality, as reflected by an increased BER. 
    Figures~\ref{fig:Motivation}(b-c) further visualize how these approaches distort ideal signals and cause high BERs. More details on this are presented in later sections.
    \item {\bf Payload bit-sequence dependence:}
    Many existing approaches assume prior knowledge of the exact signal portion (e.g., the preamble) used by the victim for fingerprint extraction, and craft their attacks by modifying only that portion.   
    However, practical RFFP systems may rely on dynamically varying payload bit sequences. Therefore, impersonation attacks must decouple the transmitted signal content from the hardware signature, so as to preserve the target's RF fingerprint across any arbitrary bit sequences. 
    
    %
     \item {\bf Unrealistic access model:}
     Most prior works assume full access to the victim classifier. This is unrealistic in practice, as such DL models are typically inaccessible. Moreover, these attacks are often evaluated via direct signal injection rather than via over-the-air (OTA) transmissions, thereby bypassing the complexities and distortions introduced by the wireless channel and the transmitter hardware.
\end{enumerate}

\subsection{Main Contributions}
This paper proposes a realistic impersonation framework that addresses the aforementioned limitations of prior works. Our key contributions can be summarized as follows:
\begin{itemize}[noitemsep, topsep=0pt, leftmargin=*]
    \item \textbf{Impairment-Driven Signal Generation Framework:} 
    We introduce a novel DL-based framework that explicitly estimates and applies hardware impairments to generate impersonation signals with controllable bit sequences, all without requiring any prior knowledge of the victim RFFP classification model. 
    To the best of our knowledge, this is the first framework capable of operating under unknown target models and system conditions.

    \item \textbf{Robust Signal Impersonation Under Domain Shift:} 
    We demonstrate the effectiveness of our impersonation attack framework in terms of its generalizability across different domains (i.e., receivers, wireless/wired, locations, channels) and its robustness under different domain-adaptive data representations.
    To the best of our knowledge, this work is the first to study the robustness of RFFP impersonation attacks against changing domains.

\item \textbf{Bluetooth RFFP Dataset Release:}
Comprehensive~\href{https://research.engr.oregonstate.edu/hamdaoui/sites/research.engr.oregonstate.edu.hamdaoui/files/release_note_datasets_ble_august2025_v1.pdf}{{\textcolor{blue}{BLE RFFP Datasets}}} collected from 31 IoT devices across various environments, receivers, and channels are made publicly available. These datasets are valuable for studying RFFP domain generalization, attacks, and impairment estimation.

    
\end{itemize}

The remainder of this paper is organized as follows.
Section~\ref{sec:taxonomy} presents our threat model.
Section~\ref{sec:method} presents the proposed framework.
Section~\ref{sec:unsupervisedlearning} introduces the unsupervised learning-based impairments estimation.
Section~\ref{sec:dataset} describes the testbed and datasets.
Sections~\ref{sec:results} and~\ref{sec:results2} present the results under direct injection and OTA setups, respectively.
Section~\ref{sec:conclusion} concludes the paper.


\section{Adversarial Taxonomy and Threat Model}
\label{sec:taxonomy}


Adversarial machine learning traditionally considers three threat models~\cite{costa2024deep}: \textit{White Box} (full access to model parameters, gradients, and output logits), \textit{Black Box} (access only to the output logits), and \textit{Gray Box} (partial model knowledge). However, these categories alone are insufficient to fully characterize an RFFP attacker's capability. 

\subsection{RFFP Attack Access Capability}

\begin{figure}
    \centering
    \setlength{\subfigcapskip}{-7pt}
    \subfigure[Full Access]{\includegraphics[width=0.325\linewidth]{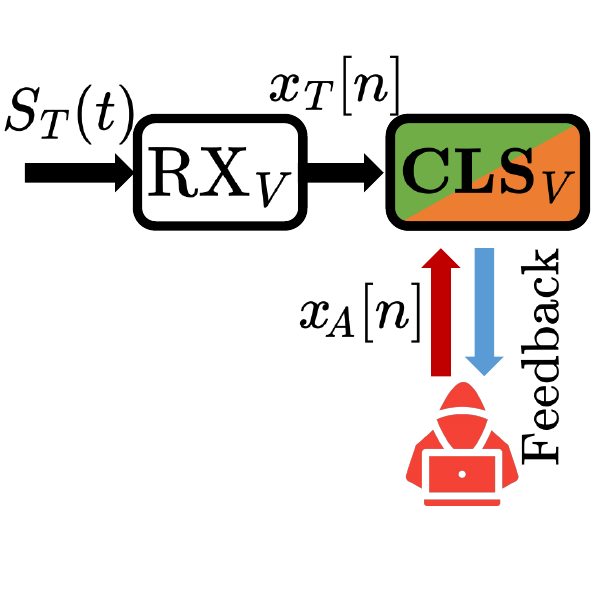}}
    \subfigure[Limited Access]{\includegraphics[width=0.325\linewidth]{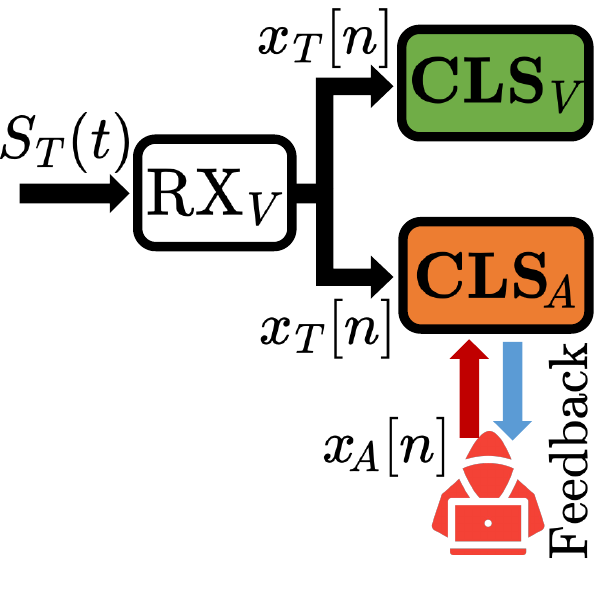}}
    \subfigure[No Access]{\includegraphics[width=0.325\linewidth]{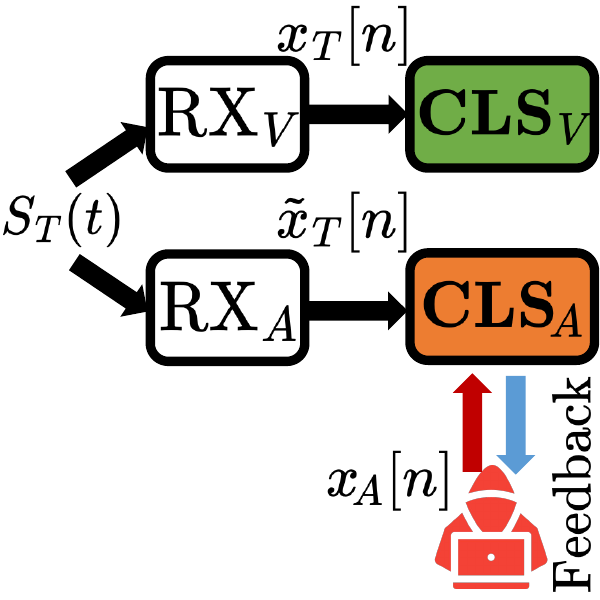}}
    \caption{
    Threat model anatomy in RFFP attacks.}
    \label{fig:AccessAssumptions}
\end{figure}

Additional access capability emerges in RFFP systems due to their physical nature. Specifically, the access threat model naturally leads to three practical categories (shown in Figure~\ref{fig:AccessAssumptions}):
\begin{enumerate} [noitemsep, topsep=0pt, leftmargin=*]
    \item \textit{Full-Access}: The attacker has access to both the victim's receiver hardware (RX$_V$) and its RFFP classifier ($\mathbf{CLS}_V$). 

    \item \textit{Limited-Access}: The attacker has no access to $\mathbf{CLS}_V$, but controls RX$_V$, and thus can collect data to train a local surrogate classifier $\mathbf{CLS}_A$ that mimics the victim's $\mathbf{CLS}_V$. 
    
    \item \textit{No-Access}: 
    This corresponds to the most realistic scenario in which the attacker uses its own receiver RX$_A$ to sample devices' transmitted signals and uses the sampled data to train a local classifier $\mathbf{CLS}_A$ to mimic a target $\mathbf{CLS}_V$. 
     
\end{enumerate}

It is important to mention that the \textit{No-Access} setting introduces additional challenges: signals captured through a receiver with different hardware and from a different location experience distinct channel conditions and embed additional receiver impairments, making it significantly harder to train a surrogate that faithfully approximates the target classifier.  

\subsection{RFFP Attack Launching Capability}
\begin{figure}
    \centering
    \setlength{\subfigcapskip}{-4pt}
    \subfigure[Direct Injection Attack]{\includegraphics[width=0.45\linewidth]{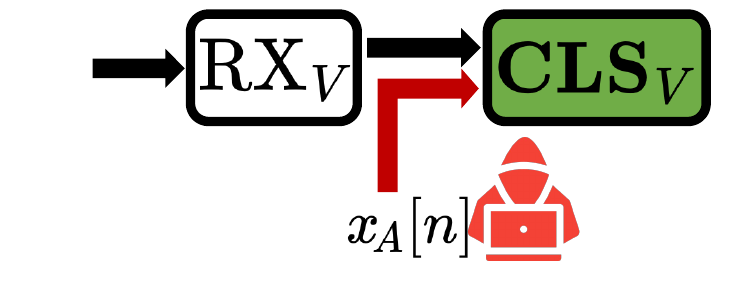}}
    \subfigure[Over-the-Air Attack]{\includegraphics[width=0.45\linewidth]{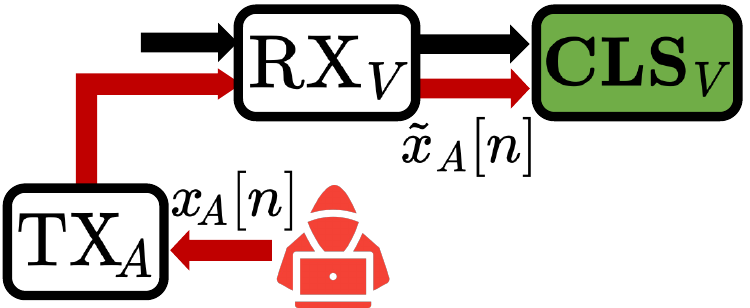}}
    \caption{RFFP attack deployment strategies.} 
    \label{fig:AttackAssumptions}
\end{figure}

Once the attack signal, $x_A[n]$, is crafted by the attacker, the attacker's goal is then to feed it to the victim classifier ($\mathbf{CLS}_V$) and make the victim device (running $\mathbf{CLS}_V$) mistakenly classify it as the targeted device.
To validate these attacks, two strategies can be considered (shown in Figure~\ref{fig:AttackAssumptions}):
\begin{itemize} [noitemsep, topsep=0pt, leftmargin=*]
\item \textit{Direct Injection}: The attacker's crafted signal (samples) is directly fed into the victim's classifier without physical transmission, an assumption adopted by most prior studies.

\item \textit{Over-the-Air (OTA)}: The attacker transmits the crafted signal using its own transceiver, allowing it to propagate through a different wireless channel before being received and processed by the victim classifier $\mathbf{CLS}_V$.
While more realistic, this attacking strategy is more challenging to launch. This is because the signal received by the victim undergoes multiple distortions, including  impairments introduced by the attacker's transmitter, channel effects during propagation, and impairments from the victim's receiver.

\end{itemize}

In this work, we test and validate the proposed attack framework using both strategies, Direct Injection and OTA. For this, we assume that the attacker has an independent transceiver (RX$_A$), capable of capturing and transmitting signals over the air. In our work, we focus on the \textit{No Access} model, where the attacker has no access to the victim's receiver (RX$_V$) nor its classifier ($\mathbf{CLS}_V$), but can instead capture signals from the targeted device through its own hardware RX$_A$.

\section{Targeted Impersonation Attack Framework}
\label{sec:method}
Our proposed framework consists of two variants. The first employs a trainable impairments estimator, termed \textbf{HWE}, to predict the target device impairments and embed them into a maliciously crafted payload, thereby synthesizing a physically consistent impersonation signal. The second variant, termed \textbf{HWE+}, builds upon \textbf{HWE} by introducing an additional gradient-based refinement stage to enhance the impersonation signal. These approaches are detailed in the following sections; before presenting them, however, we first provide background on hardware impairments to establish the necessary context.

\subsection{Hardware Impairments Background}
\begin{figure*}
    \centering
    \setlength{\subfigcapskip}{-5pt}
    \includegraphics[width=\linewidth]{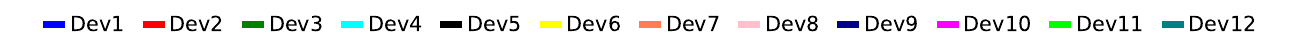}
    \vspace{-22pt}
    
    \subfigure[RX$_V$-Ch1 (wired)]{\includegraphics[width=0.2\linewidth]{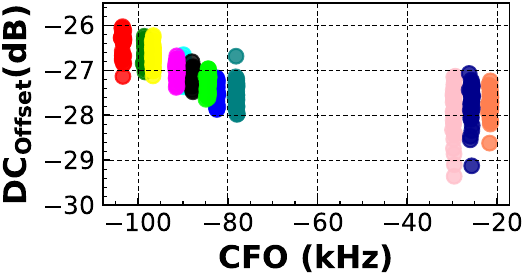}}\label{subfig:cfo}
    \hspace{-7pt}
    \subfigure[RX$_A$-Ch1 (wired)]{\includegraphics[width=0.2\linewidth]{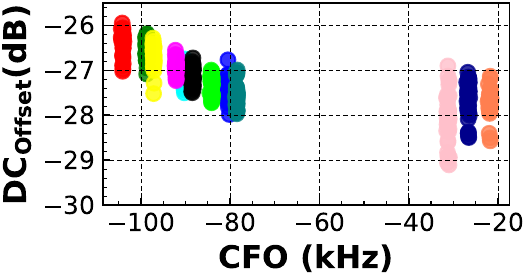}}
    \hspace{-7pt}
    \subfigure[RX$_A$-Ch2 (wired)]{\includegraphics[width=0.2\linewidth]{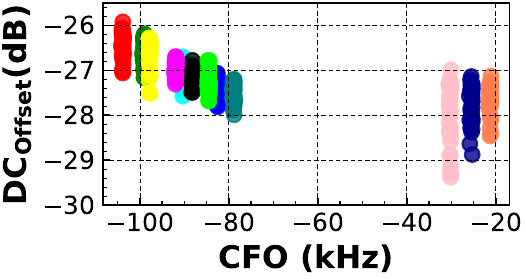}}
    \hspace{-7pt}
    \subfigure[RX$_A$-Ch14 (wired)]{\includegraphics[width=0.2\linewidth]{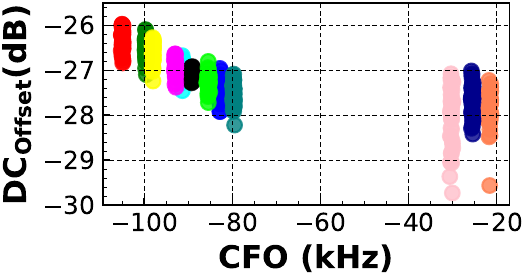}}
    \hspace{-7pt}
    \subfigure[RX$_A$-Ch1 (wireless)]{\includegraphics[width=0.2\linewidth]{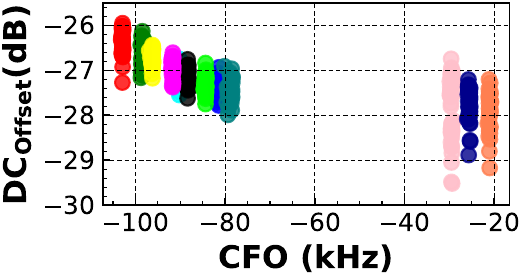}}
    \caption{Estimated DC$_{\text{Offset}}$ vs CFO: each point represents the average of ten independent estimates.}
    \label{fig:Impairments-vis}
    \vspace{-16pt}
\end{figure*}

Hardware impairments are intrinsic imperfections introduced during the manufacturing of RF components, such as local oscillators, mixers, and power amplifiers (PA), that cause transmitted signals to deviate from their ideal characteristics. In this work, we focus on the following impairments:
\begin{itemize}[noitemsep, topsep=0pt, leftmargin=*]
\item \textbf{Carrier Frequency Offset ($f_{{CFO}}$) and Phase Offset ($\theta_{{PO}}$):}
Carrier frequency offset (CFO) arises from frequency mismatches between the transmitter and receiver local oscillators, resulting in a frequency shift and a progressive phase rotation over time. In contrast, phase offset (PO) originates from phase mismatches and manifests as a constant phase shift. Both impairments stem from inherent imperfections in the local oscillator and phase-locked loop circuitry.

\item \textbf{Distorted Bandwidth-Time ($\widetilde{BT}$) Product:} This is a deviation in the Gaussian filter bandwidth-time product that is used in BLE GFSK modulation to control the pulse shaping smoothness. This affects the transient and spectral characteristics of the signal.

    \item \textbf{Maximum Frequency Deviation Offset ($\Delta f$):} 
    This represents the error in the nominal frequency deviation, that is defined as the difference between the maximum positive frequency and the central frequency in GFSK modulation.

    \item \textbf{I/Q Amplitude/Phase Imbalances ($IQ_{{Amp}}$ and $IQ_{{Phase}}$):} 
    These capture the mismatch between the in-phase (I) and quadrature (Q) components of the transmitter and receiver in amplitude/phase, most often due to imperfections in mixers and local oscillators. These typically result in constellation distortion and image frequency leakage.
    
    \item \textbf{I/Q DC Offset ($I_{{DC}}$ and $Q_{{DC}}$):} These occur when the I/Q signal origin shifts from zero due to mixer leakage or DAC imperfections, producing a constant DC component and carrier feedthrough in the transmitted signal.
\end{itemize}
Despite extensive engineering efforts to mitigate hardware impairments as typically are considered undesirable in conventional communication systems, they cannot be entirely eliminated. As a result, residual distortions persist in the received RF signal, which can serve as distinctive device fingerprints that can be exploited to uniquely identify devices. 
To illustrate, we show in Figure~\ref{fig:Impairments-vis}
the DC offset (defined as $10 \log (I_{DC}^2 + Q_{DC}^2)$) 
and CFO impairments extracted from the signals sent by 12 different BLE devices, where the same BLE signals are collected under different deployment settings (i.e., using two different receivers (RX$_V$ and RX$_A$), under three different BLE channels (Ch1, Ch2, and Ch14), and over wired and wireless links). 
The figure clearly shows that the impairments extracted from the signals---using our proposed impairments estimation approach presented later---are (i) device-specific (distinguishable across devices) and (ii) domain-agnostic (insensitive to deployment change), making them strong signatures that can be used for device identification and authentication.
Throughout, ${\Theta}$ will be used to denote the vector comprising the eight described impairments; i.e., $ {\Theta} = (\widetilde{BT}, \Delta f, f_{CFO}, \theta_{PO}, I_{DC}, $ $ Q_{DC}, IQ_{Phase}, IQ_{Amp}) $.

\subsection{\textbf{HWE}: Estimation-Based Impersonation Attack Framework} 
\begin{figure}
    \centering
    \setlength{\subfigcapskip}{-10pt}
    \subfigure[HWE]{\includegraphics[width=0.50\linewidth]{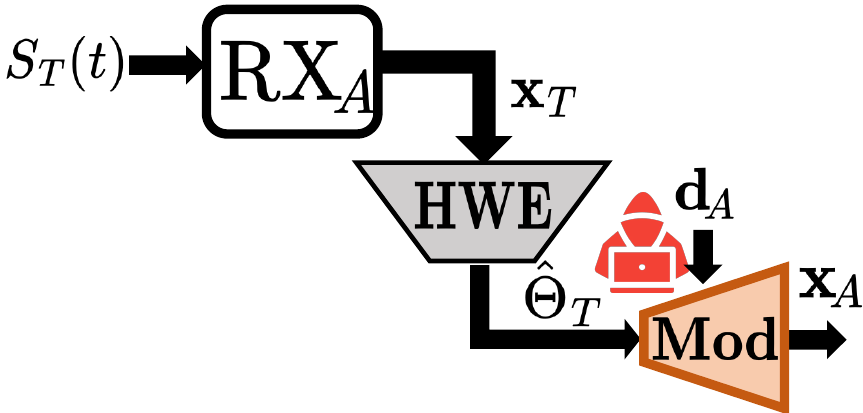}\label{subfig:HWE}}
    \hspace{-7pt}
    \subfigure[HWE+]{\includegraphics[width=0.50\linewidth]{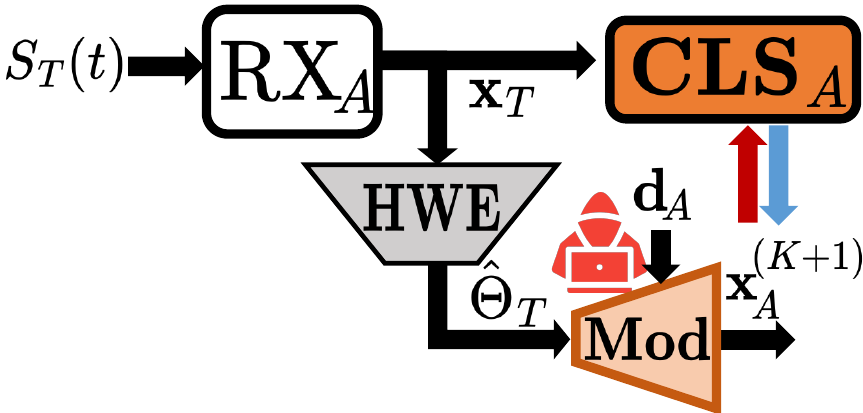}\label{subfig:HWE+}}
    \vspace{-5pt}
    \caption{Proposed impersonation attack framework.}
    \label{fig:HWEandHWE+}
\end{figure}

Figure~\ref{subfig:HWE} illustrates the overall workflow of our impersonation attack framework. The attack begins by capturing an over-the-air signal \( x_T[n] \) transmitted by the target device and estimating the device's hardware impairment vector $\hat{\Theta}_T$ through the Hardware Estimation (\textbf{HWE}) block. 
\textbf{HWE} implements a mapping function that extracts and returns the device's hardware impairments vector $\hat{\Theta}_T$ from any of the device's captured signals $x_T[n]$; i.e., $\hat{\Theta}_T = \mathbf{HWE}(x_T[n])$. The details of the proposed $\mathbf{HWE}$ function estimator are provided in Section~\ref{sec:unsupervisedlearning}.
We further define a modulation (signal generation) function $\mathbf{Mod}(\cdot, \cdot)$ at the attacker side. This function takes any impairment vector and embeds it into a controllable digital bit sequence.  
Therefore, once the target impairment vector $\hat{\Theta}_T$ is estimated, the attacker can embed it into a malicious payload $d_A[n]$ using the $\mathbf{Mod}(\cdot, \cdot)$ function, producing the synthesized impersonation signal
$x_A[n] = \mathbf{Mod}(d_A[n], $ $\hat{\Theta}_T)$.
The details of the $\mathbf{Mod(\cdot,\cdot)}$ function block are provided in Section~\ref{subsec:mod-ble}.

\subsection{\textbf{HWE+}: Gradient-Based Attack Refinement}
We extend \textbf{HWE} by leveraging gradient information from the victim classifier to further refine the attack signal. We refer to this extended method as \textbf{HWE+}. 
As shown in Figure~\ref{subfig:HWE+}, the first step remains the same, that is generating a coarse attack signal $x_A[n]$ through the \textbf{HWE} block. While this signal already embeds the target device's impairments, it may not perfectly match the decision boundaries of the victim classifier that the attacker seeks to access. 
For that, the attacker could leverage gradient information of the victim's classifier $\mathbf{CLS}_{V}(\cdot)$ to shift the distribution of the attack signal toward the target decision boundaries.
However, as having access to $\mathbf{CLS}_{V}(\cdot)$ is not realistic (as in \textit{No-Access} threat model), we propose training a local surrogate classifier $\mathbf{CLS}_{A}(\cdot)$ at the attacker side to approximate the behavior of $\mathbf{CLS}_{V}(\cdot)$.
Once $\mathbf{CLS}_{A}(\cdot)$ is fully trained (see Appendix~\ref{apn:CLSs}), we use an iterative refinement approach inspired by the PGD algorithm~\cite{kurakin2018adversarial} to maximize the likelihood of impersonating a target device.

To formalize our refinement process, let $\mathcal{L}(\cdot,\cdot)$ be the loss function (e.g., cross-entropy) used to train $\mathbf{CLS}_{A}(\cdot)$ and $y_T$ be the label of the target device. 
The attacker objective is to add small perturbation $\mathbf{\delta}\in \mathbb{C}^N$ to the synthesized attack signal $\mathbf{x}_A=\left[x_A[0], \dots, x_A[N\!-\!1]\right]$ $\in $ $\mathbb{C}^N$ such that the classifier is more likely to output 
$y_T$. This is equivalent to solving:
$$\min_\delta \ \mathcal{L}(\mathbf{CLS}_{A}(\mathbf{x}_A+\mathbf{\delta}), y_T) \quad \text{s.t.} \quad \|\delta\|_{\infty} \leq \epsilon$$
where $\epsilon$ is the $L_{\infty}$ perturbation budget, i.e., the maximum allowable magnitude of the adversarial modification applied to any individual signal sample.
Starting from the initial signal $\mathbf{x}_A^{(0)}=\mathbf{x}_A$, the attacker refines the attack signal iteratively using PGD~\cite{kurakin2018adversarial}, where the $K$-th iterative update is given by:
\begin{equation} 
\!\!\! \mathbf{x}_A^{(K+1)} \!\!\!=\! \text{Clip}^{\mathbf{x}_A}_{ \epsilon}\!\!\left\{\mathbf{x}_A^{(K)} \!\!-\! \alpha \ \! \text{sign} (\nabla_\mathbf{x} \mathcal{L}(\mathbf{CLS}_{A}(\mathbf{x}_A^{(K)}), y_T)) \!\right\}
\label{eq:PGD}
\end{equation}
where $\text{Clip}^{\mathbf{x}_A}_{\epsilon}(\mathbf{z})= \min\left(\max(\mathbf{z},\mathbf{x}_A-\epsilon), \mathbf{x}_A +\epsilon\right)$  is a clipping function that ensures the perturbation remains within the valid bounded region with a radius $\epsilon$ and centered at $\mathbf{x}_A$. $\alpha$ is the step size. Moreover, we use 200 update steps with $\epsilon=0.05$ and $\alpha=0.001$, ensuring no increase in BER.
 

\section{Learning-Based Impairments Estimation}
\label{sec:unsupervisedlearning}
Our attack framework centers on the hardware impairments estimation block (\textbf{HWE}), which implements a mapping function that extracts and returns the device-specific impairment vector $\hat{\Theta}$ from any of the device's captured signal $x[n]$. This mapping can be realized using various approaches, including iterative optimization techniques~\cite{givehchian2022evaluating}, moment-matching estimators~\cite{kay1993fundamentals}, or a neural network (NN) models. 
Since accurate estimation of the device's hardware impairments is critical to our proposed impersonation attack, a precise and tunable estimation approach is essential. We therefore adopt an NN-based implementation and train a dedicated neural network estimator $\mathbf{HWE}(\cdot;\Phi)$ parameterized by $\Phi$. This choice is justified through the experimental results presented in Section~\ref{subsec:hwe-eval} and later in Sections~\ref{sec:results} and~\ref{sec:results2}.
Training the estimator block in a supervised manner requires access to the impairments' ground-truth values---very expensive and often infeasible task. To address this, we propose a novel unsupervised learning framework that enables end-to-end training of the $\mathbf{HWE}(\cdot;\Phi)$ block, allowing it to accurately estimate device-specific impairments without requiring ground-truth impairment labels.

The training approach of our $\mathbf{HWE}(\cdot;\Phi)$ estimator, which is shown in Figure~\ref{fig:HW blocks}, employs an encoder-decoder architecture, where the trainable $\mathbf{HWE}(\cdot;\Phi)$ maps (encodes) any input signal to its corresponding impairment vector $\hat{\Theta}$, and the non-trainable generator (modulation) function $\mathbf{Mod}(\cdot,\cdot)$ reconstructs (decodes) the signal given an arbitrary impairment vector $\hat{\Theta}$.
Once trained, the resulting block $\mathbf{HWE}(\cdot;\Phi)$ is integrated into the targeted impersonation attack framework presented in Section~\ref{sec:method}. 
We now provide detailed description of the training procedure along with descriptions of the other blocks constituting the framework shown in Figure~\ref{fig:HW blocks}.


\subsection{Unsupervised Learning Training Process}
Let $x[n]$ be any captured signal whose impairments we aim to estimate and $d[n] = \mathbf{Dem}(x[n])$ be a demodulation function that takes $x[n]$ and outputs its underlying bit sequence $d[n]$. Let $x_S[n;\Theta] = \mathbf{Mod}(d[n], \Theta)$ be a modulation function that takes a controllable bit sequence $d[n]$ and an impairments vector ${\Theta}$, to generate the synthetic signal $x_S[n;\Theta]$. These $\mathbf{Mod}(\cdot, \cdot)$ and $ \mathbf{Dem}(\cdot)$ functions are described in Sections~\ref{subsec:mod-ble}  and~\ref{subsec:ble-demod}. 
The goal is then to train  $\mathbf{HWE}(\cdot;\Phi)$, parameterized by learnable weights $\Phi$, for the task of mapping signal $x[n]$ to its underlying impairments vector $\hat{\Theta}$. 

Through our experimentation, we found that using the phase of the signal $x[n]$ as an input, i.e., $\hat{\Theta} = \mathbf{HWE}(\angle_ux[n];\Phi)$, yields more precise impairment estimates, and was therefore adopted for training and inference. This is because the phase information directly reflects the hardware-induced impairments, such as carrier frequency offset (CFO), phase offset, and I/Q imbalance, much better than the amplitude. 
\begin{figure}
    \centering
    \includegraphics[width=0.8\linewidth]{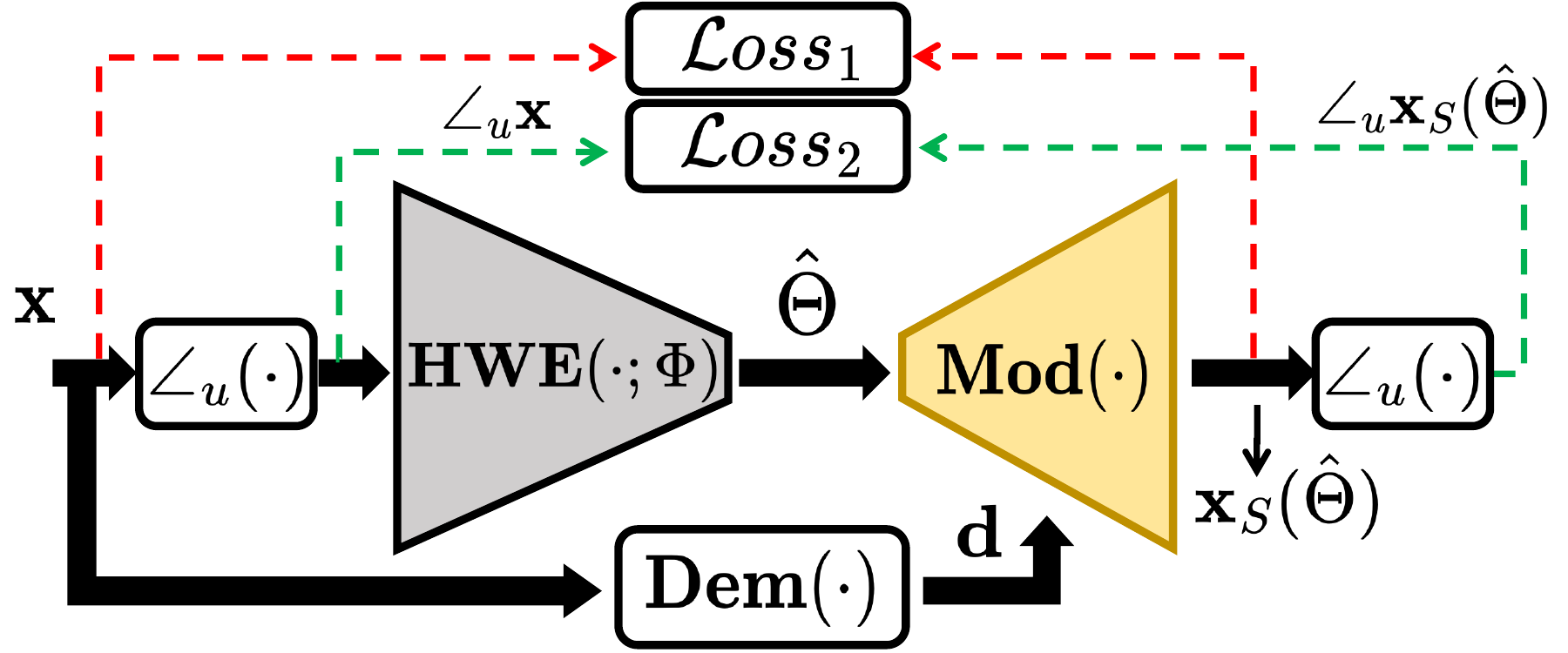}
    \caption{The proposed NN-based training framework.}
    \label{fig:HW blocks}
\end{figure}
To this end, we propose a novel unsupervised learning framework for $\mathbf{HWE}(\cdot;\Phi)$ training. As shown in Figure~\ref{fig:HW blocks}, our approach follows the Encoder-Decoder architecture, where the $\mathbf{HWE}(\cdot;\Phi)$ serves as the trainable encoder (compressor) producing the latent hardware impairment vector with the input signal, and $\mathbf{Mod}(\cdot,\cdot)$ functions as the decoder (generator) that reconstructs the signal from the estimated impairments.
Formally, once the signal $\mathbf{x} = \left[x[0], \dots, x[N\!-\!1]\right]\in \mathbb{C}^N$ is captured by the attacker hardware, the impairments vector is extracted through $\hat\Theta = \mathbf{HWE}(\angle_u\mathbf{x};\Phi)$ and the bit sequence associated with $\mathbf{x}$ is extracted through the $\mathbf{Dem}(\cdot)$ block.
This bit sequence, together with the estimated impairments $\hat{\Theta}$, is fed into the generator block ${\mathbf{Mod}}\left( \cdot \, , \cdot  \right)$ to reconstruct the signal $\mathbf{x}_S(\hat{\Theta}) = \left[x_S[0,\hat{\Theta}], \dots, x_S[N\!-\!1,\hat{\Theta}]\!\right]\in \mathbb{C}^N$.
Our training thus finds the optimal NN parameters $\Phi$ such that the latent impairment vector $\hat\Theta$ perfectly aligns  $\mathbf{x}$ and $\mathbf{x}_S(\Theta)$ signals. We used the reconstruction loss between the captured signal $\mathbf{x}$ and the reconstructed version $\mathbf{x}_S$ as our training objective, which is defined as 
$\min_{\Phi} \ \left\|\mathbf{x}_S(\hat\Theta _{\Phi}) - \mathbf{x} \right\|_p$,
where $\| \cdot\|_p$ represent the $p$-norm. We further enhance this loss by incorporating a regularized phase-alignment reconstruction loss--following the approach  in~\cite{albousayri2025neural}. Accordingly, the objective becomes: 
\begin{equation} \label{eq:phimin2}
\min_{\Phi} \ \left\|\mathbf{x}_S(\hat\Theta _{\Phi}) - \mathbf{x} \right\|_p +\lambda_{\text{Phase}} \left\| \angle_u\mathbf{x}_S(\hat\Theta _{\Phi}) - \angle_u\mathbf{x} \right\|_p,
\end{equation}
where $\angle_u(\cdot)$ represents the unwrapped phase and $\lambda_{\text{Phase}}$ is a regularization factor that balances IQ loss and phase alignment loss. The phase alignment loss transforms the representation into a linear function of key parameters such as CFO and PO~\cite{mckilliam2010frequency}. These two key impairments, not only important for RFFP identification, but also necessary for the accurate estimates of the remaining impairments. For that, we propose using a decaying scheduler for $\lambda_{\text{Phase}}$ starting from $\lambda^{\text{max}}_{\text{Phase}}$ and ending at $\lambda^{\text{min}}_{\text{Phase}}$, so that the weight of the phase loss term starts of high and gradually decays for better refinement of impairments values at later stages. 

Overall, unsupervised training is made possible in this setting because Equation \eqref{eq:phimin2} can be written as a function of the unlabeled signals $\mathbf{x}$ through $\textbf{HWE} (\cdot; \Phi)$. In other words, given $\mathbf{x}_S(\hat\Theta)=\mathbf{Mod}(\mathbf{d}, \hat\Theta)= \mathbf{Mod}(\mathbf{Dem}( \mathbf{x}), \hat\Theta)$,
the estimator can be trained by minimizing the following loss:
\begin{align} \notag
\min_{\Phi}& \ \| \mathbf{Mod}(\mathbf{Dem}( \mathbf{x}), \textbf{HWE} (\angle_u\mathbf{x}; \Phi)) - \mathbf{x} \|_p + \\ &  \!\!\!\! \lambda_{\text{Phase}}\| \angle_u\mathbf{Mod}(\mathbf{Dem}( \mathbf{x}), \textbf{HWE} (\angle_u\mathbf{x}; \Phi)) - \angle_u\mathbf{x} \|_p.
\label{eq: Minimization_Reg} 
\end{align}
The architecture and hyperparameters used to train $\mathbf{HWE}(\cdot;\Phi)$ are described in Appendix~\ref{apn:HWEblock}.

\subsection{The $\mathbf{Mod}(\cdot,\cdot)$ Function Block} \label{subsec:mod-ble}
Let $h[n] \!=\! (\sqrt{\pi}/{a}) \exp({- {\pi ^2 n^2}/{a^2})}$ be the impulse response of a Gaussian pulse-shaping filter where $a = ({1}/{\widetilde{BT}})\sqrt{{ln(2)}/{2}}$ is a parameter related to $3$-dB bandwidth.
%
Given the baseband data $d[n]$, the Gaussian filtered pulse can be expressed as $g[n] = d[n] * h[n]$, where $*$ represents the convolution operation.
The instantaneous angular shift function can then be expressed as discrete version of running integral $\phi [n] = 2 \pi \left(f_m + \Delta f  \right)\sum_{k=0}^{n} {g}[k] \ T_S$, 
where $f_{m}$, $\Delta f$ and $T_S$ represent the optimal peak frequency deviation, the offset from the optimal value of $f_m$ and the sampling interval, respectively. 

The generated baseband signal can be expressed as~\cite{givehchian2022evaluating} 
${x}_S[n] = \left(\tilde{x}_I [n]  +  j\tilde{x}_Q  [n]\right)e^{j\left(2\pi f_\text{CFO}  n T_s+ \theta_{PO}\right)}$ with $\tilde{x}_I[n] = (1-IQ_{Amp})\cos(\phi[n] - {IQ_{Phase}}/{2}) + I_{DC}$ and
$\tilde{x}_Q[n] = (1+IQ_{Amp})\sin(\phi[n] + {IQ_{Phase}}/{2}) +Q_{DC}$. 
Since the signal $x_{S}[n]$ depends on the impairments vector $ {\Theta} = (\widetilde{BT}, \Delta f, f_{CFO}, \theta_{PO}, I_{DC}, $ $ Q_{DC}, IQ_{Phase}, IQ_{Amp})$,  $\mathbf{Mod}(d[n], {\Theta})$ can  be implemented as 
\( x_{S}[n; {\Theta}] \triangleq x_{S}[n]\).

\subsection{The $\mathbf{Dem}(\cdot)$ Function Block}\label{subsec:ble-demod}
Demodulation, the process of retrieving the baseband bit sequence from a received signal, is essentially the inverse of the signal modulation steps described earlier. While demodulating a signal under high levels of impairment-induced distortion can lead to errors, it remains feasible when the impairments are sufficiently small. 
In BLE, the original bit sequence can be recovered by computing the discrete-time derivative  $\Delta_{n}\angle_u{x[n]}$ of the unwrapped phase of $x[n]$. 



\subsection{Evaluation of the Estimation Framework}\label{subsec:hwe-eval}
This section evaluates the proposed impairments estimation approach (once fully trained) against established baseline methods, with primary focus on CFO. 
We consider two key aspects: (i) the estimation methods' ability to accurately reconstruct the captured signal, and (ii) their computational efficiency, measured by the time required to estimate the impairments.
%
%
The baseline approaches operate on the preamble portion of the captured signal and are widely adopted in industry due to their lightweight implementation and computational efficiency. These methods are summarized below.
\begin{enumerate} [noitemsep, topsep=0pt, leftmargin=*]
\item \textbf{Correlation-based ($\tt{MOM}$)~\cite{kay1993fundamentals}:} It estimates CFO using the autocorrelation of the captured signal. Specifically, it measures the phase difference between the signal and a one-sample-delayed version of itself, which directly reflects the frequency offset.

\item \textbf{Extended Correlation-based ($\tt{MOM+}$)~\cite{kay1993fundamentals}:} This extends the basic correlation estimator by averaging phase differences across multiple time shifts to improve accuracy.

\item \textbf{Mean Phase Derivative ($\tt{MPD}$)~\cite{mckilliam2010frequency}:} This approach estimates CFO by averaging the periodic phase derivative between consecutive samples of the signal.

\item \textbf{Gradient Descent ($\tt{GD}$)~\cite{givehchian2022evaluating}:} This jointly estimates the set of 8 impairments using an iterative gradient descent ($\tt{GD}$) optimizer to find the optimal vector of impairments that minimize the objective in Eq.~\eqref{eq:phimin2}. The stopping criterion was defined as the point at which the loss of the iterative procedure became less than or equal to the loss achieved by our proposed $\mathbf{HWE}$ method.

\end{enumerate}

\subsubsection{Signal Reconstruction Loss}

Once we capture the device's signal $\mathbf{x}$ and estimate the impairments $\hat\Theta$ using the method under study, we synthesize the signal $\mathbf{x}_S(\hat\Theta)$ as described in Section~\ref{subsec:mod-ble}. We then compute the mean squared error (MSE) between the original captured signal $\mathbf{x}$ and the synthesized signal $\mathbf{x}_S(\hat\Theta)$ to quantify the reconstruction loss 
$\mathbf{MSE}_{Rec} (\mathbf{x}, \mathbf{x}_S(\hat\Theta)) = \frac{1}{N} \| \mathbf{x} - \mathbf{x}_S(\hat\Theta)\|_2$ where $N$ is the signal length (number of samples). 
\begin{figure}
\centering
\includegraphics[width=0.85\linewidth]{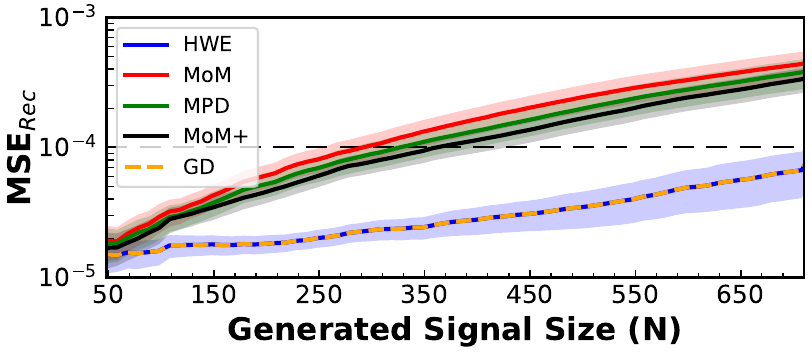}
\caption{Reconstruction loss of the studied methods.}
\label{fig:Generalization}
\end{figure}
Figure~\ref{fig:Generalization}, depicting the loss as a function of the signal length $N$, shows that as $N$ increases, our $\mathbf{HWE}$ significantly outperforms the baseline techniques. The figure also indicates that estimates of the impairments when derived from a small segment of the frame remain valid over longer signal durations, confirming that the estimated impairments capture persistent hardware characteristics rather than sequence-specific artifacts. Consequently, they enable the generation of long attack signals with controllable malicious payloads.

\subsubsection{Execution Time}
\begin{figure}
    \centering
    \includegraphics[width=0.9\linewidth]{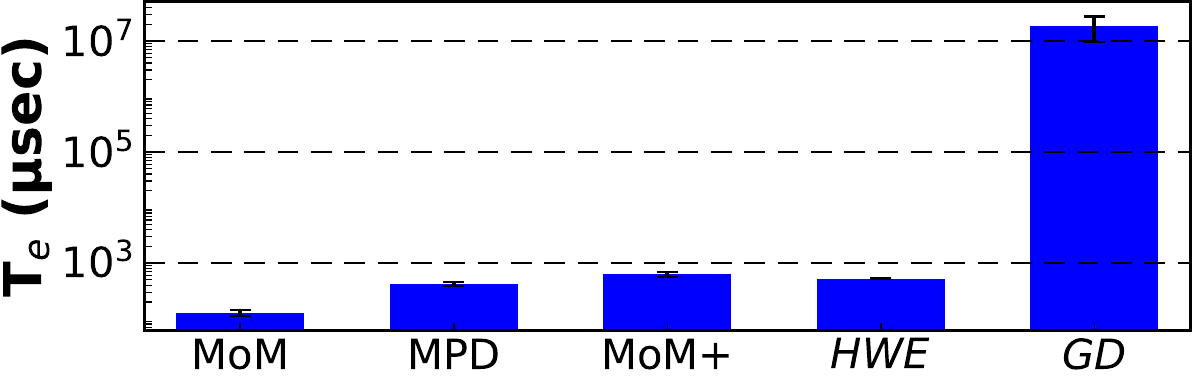}
    \caption{Execution times (log-scale) of the studied methods.}
    \label{fig:HWEvsBLsTimes}
\end{figure}

We now measure and report the execution time, $T_e$, the time it takes $\mathbf{HWE}$ and $\tt{GD}$ to estimate all eight impairments. For $\tt{MoM}$, $\tt{MoM+}$ and $\tt{MPD}$, the execution time corresponds to only the time needed for estimating a single impairment, namely CFO.
Figure~\ref{fig:HWEvsBLsTimes} presents $T_e$ for each of the considered estimation methods. The results show that, although the proposed $\mathbf{HWE}$ method requires slightly longer time to estimate $\hat{\Theta}$ compared to $\tt{MoM}$ and $\tt{MPD}$, it remains more practical for real-world deployment. This is because $\mathbf{HWE}$ jointly estimates eight hardware impairments in a single forward pass, whereas the baseline methods estimate only a single impairment at a time, requiring multiple independent executions to obtain the full impairment vector. When compared with $\tt{GD}$, our proposed estimator achieves the same reconstruction with dramatically lower cost---requiring only $0.003$ to $0.06\%$ of their average execution time---making it a much faster and more reliable solution for real-time deployment.

\section{Experimental Hardware and Setup}
\label{sec:dataset}

Our testbed (see Figure~\ref{fig:Testbed}(a)) consists of 31 Seeed Studio XIAO IoT devices---mini-development boards based on the Espressif ESP32-C3 WiFi/BLE dual-mode transmitter to send Bluetooth (BLE) traffic. We hardcoded these chips to send the signals over 3 BLE frequency channels, Channel 1 (Ch1), Channel 2 (Ch2) and Channel 14 (Ch14), with each channel being centered at 2.406GHz, 2.408GHz and 2.434GHz, respectively.  
Two Ettus Universal Software Radio Peripheral (USRP) B210 transceivers---RX1 and RX2---were employed to sample and collect the RF data from the 31 transmitters in the form of raw IQ samples via GNURadio. 
RX1 is used as the legitimate (victim) RFFP system hardware (RX$_V$), and RX2 is used as attacker's (RX$_A$) as seen in Figure~\ref{fig:Testbed}.

Before starting our data collection, each device is powered on for a 6-minute warm-up period to ensure hardware stabilization~\cite{elmaghbub2024no}, then followed by a 2-minute period of data collection. 
The bandwidth was set to 2MHz, the sampling rate was set to 6MS/s and the power gain was set to 29dB and 8dB for the wireless and wired data collections, respectively. We utilized 1M PHY modulation setup for this experiment. 

The following sections further detail the data collection process used for training the legitimate/victim RFFP system, and the attacker data spoofing stage from victim devices that try to access the legitimate/victim RFFP system. 

\subsection{Training Data Collection}  
Two data collection setups are needed, one for training the victim/legitimate classifier and one for training the attacker's $\bf HWE$ estimation framework. For the victim's classifier, we collected signals from all 31 devices under Ch1 using receiver RX$_V$ over wired transmission. This setup eliminates channel variability and mimics a secure, domain-generalized training process, as recommended in~\cite{albousayri2025bluetooth}.
For the attacker's framework training, we collected signals using receiver RX$_A$ over both wired and wireless, under different frequency channels, and at different times. For the wired setup, we used Ch1, Ch2 and Ch14. For the wireless setup, we used Ch1 while keeping $2$m separation between the IoT transmitters and the receiver.


\subsection{Adversarial OTA Setup}  \label{subsec:OTAsetup}

\begin{figure}
    \centering
    \begin{minipage}[b]{0.2\linewidth}
            \centering
            \subfigure[ESP32C3]{
                \includegraphics[width=\linewidth]{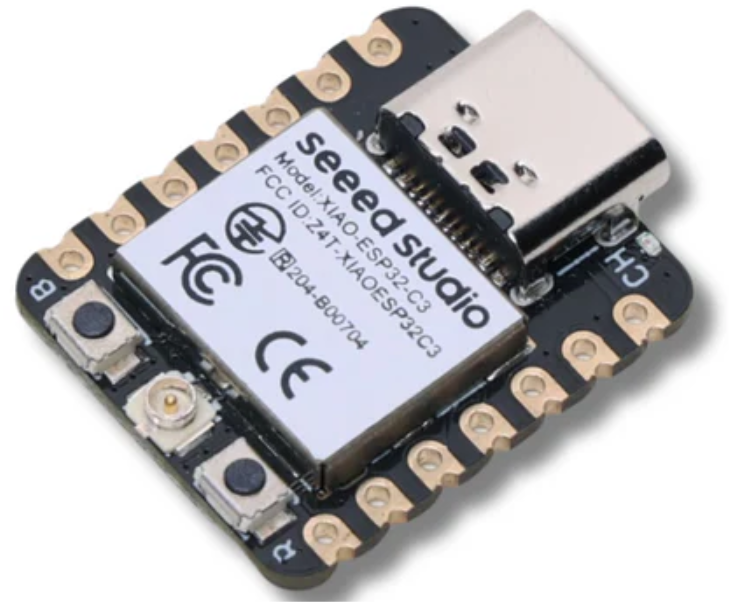}}\\[0.5ex]
                \vspace{-12pt}
            \subfigure[USRP B210]{
                \includegraphics[width=\linewidth]{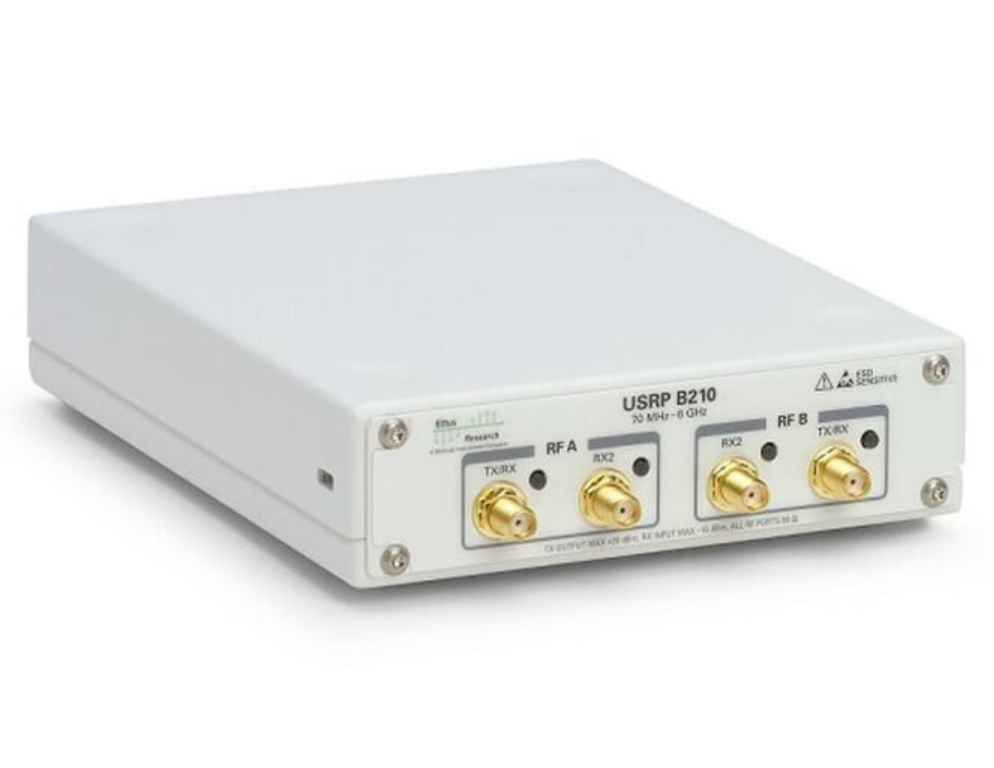}}
    \end{minipage}
    {\centering
    \begin{minipage}[b]{0.25\linewidth}
        \centering
        \subfigure[Wired]{
            \includegraphics[width=\linewidth]{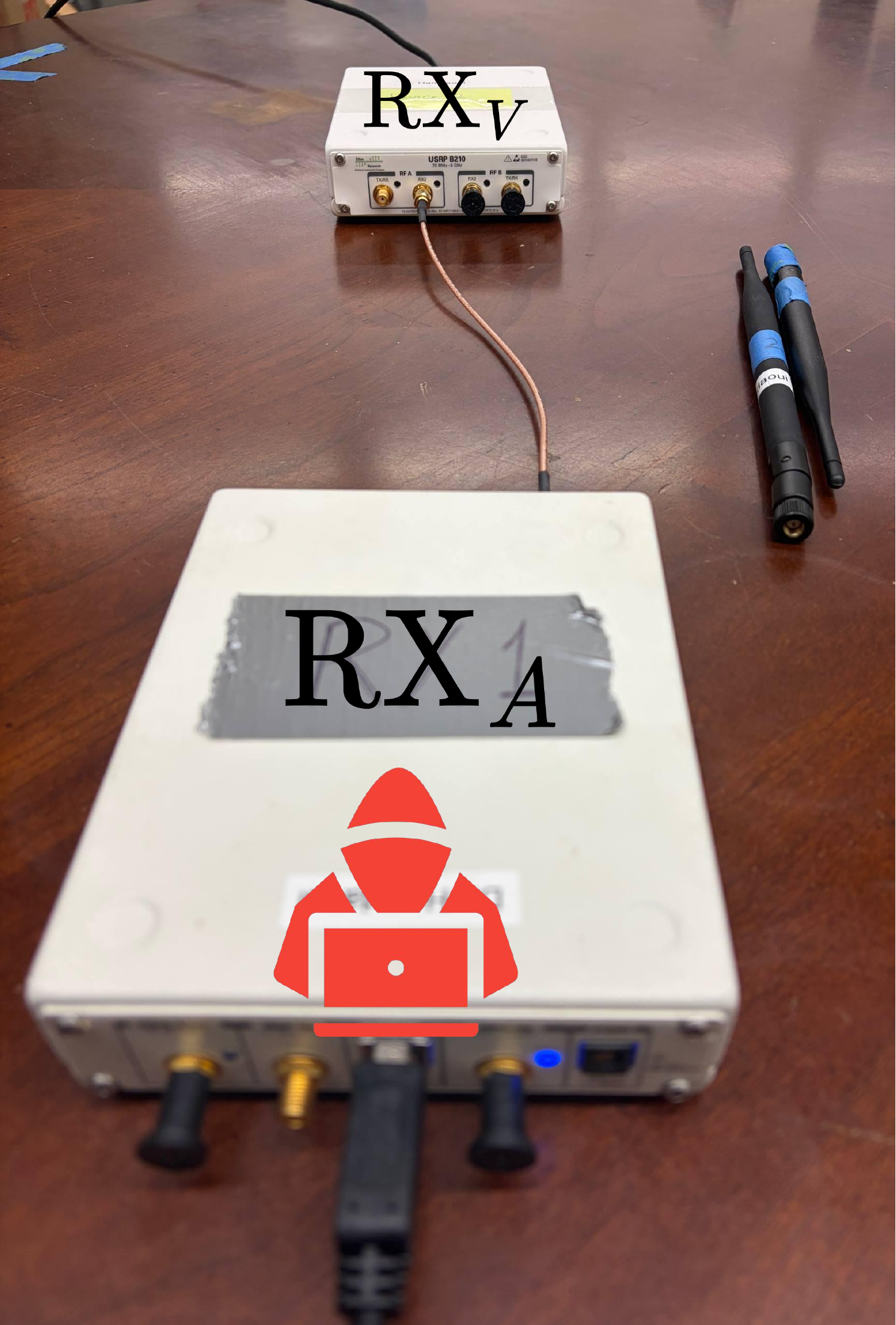}}
    \end{minipage}}
    {\centering
    \begin{minipage}[b]{0.25\linewidth}
        \centering
        \subfigure[1.5m LoS]{
            \includegraphics[width=\linewidth]{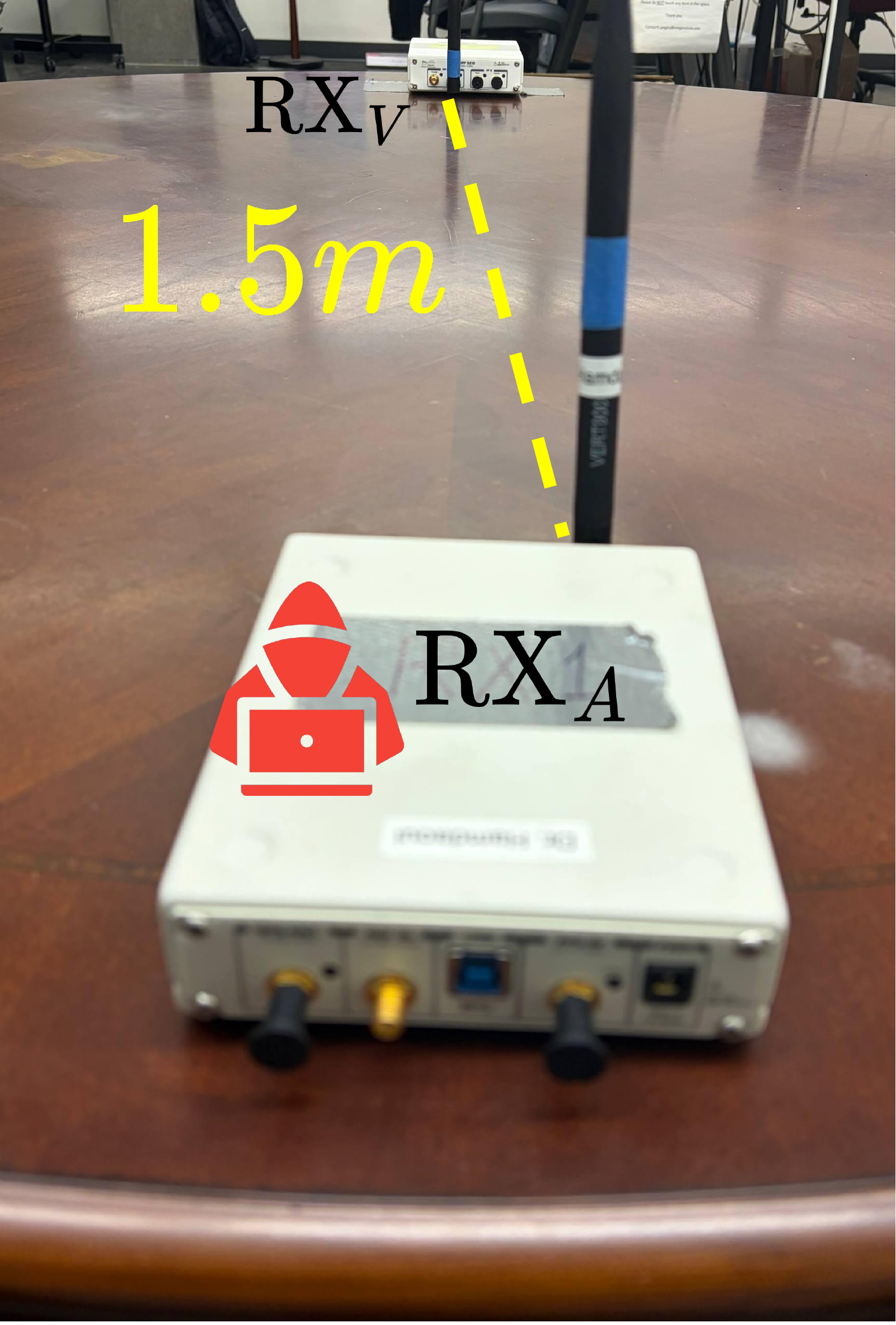}}
    \end{minipage}}
    {\begin{minipage}[b]{0.25\linewidth}
        \centering
        \subfigure[7m LoS]{
            \includegraphics[width=\linewidth]{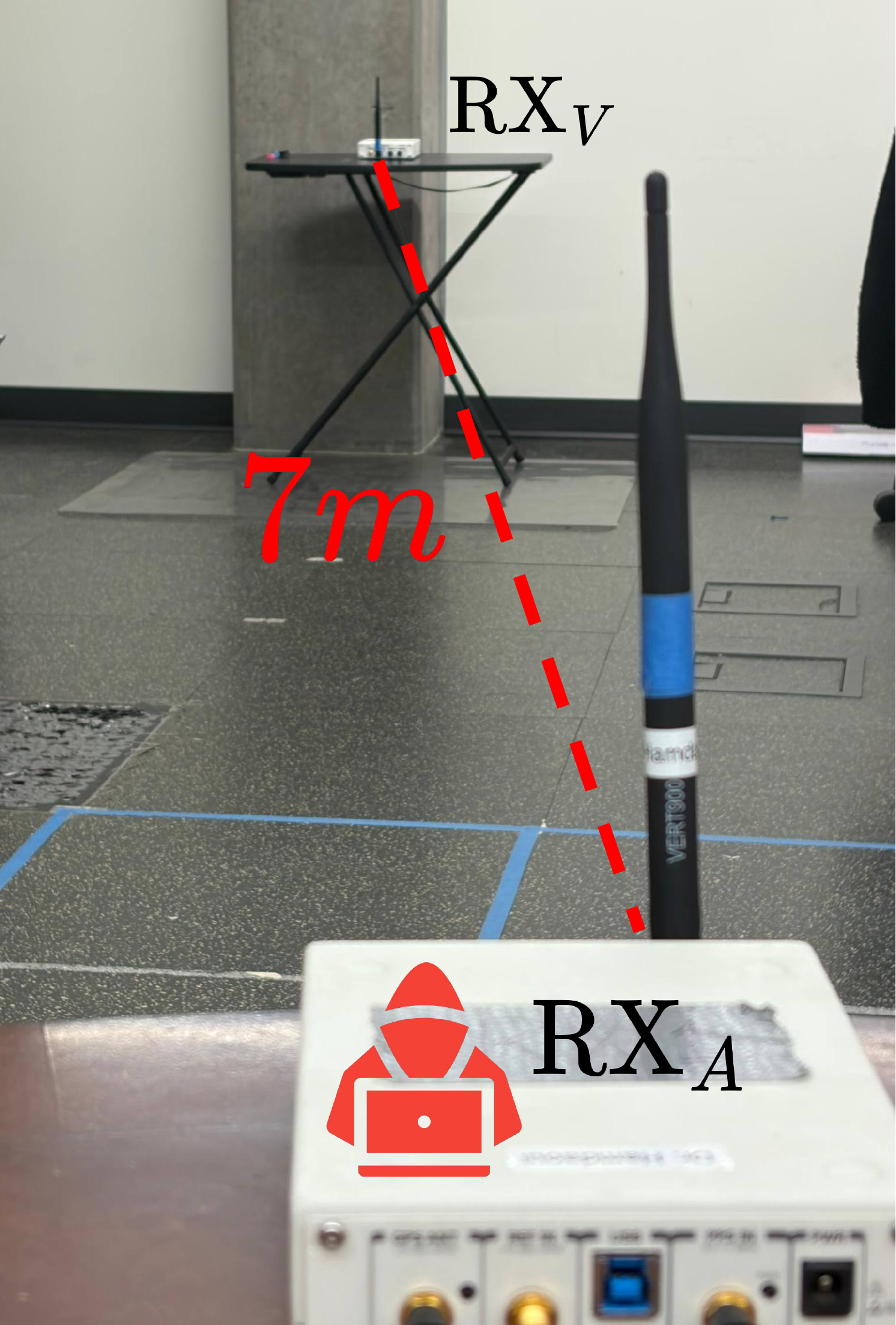}}
    \end{minipage}}
    \caption{Experimental testbed and hardware configuration.}
    \label{fig:Testbed}
\end{figure}

Once the attack signals are crafted through the proposed framework, the attacker transmits the impersonated signal to the legitimate/victim receiver RX$_V$ using its own transceiver RX$_A$ (an USRP B210) to access the legitimate RFFP system. For this impersonated transmission, we used BLE frequency channel 0 (Ch0) that is centered on 2.402 GHz carrier. 
This was done to further test the generalization of the attack, where a completely unseen channel condition is utilized. We considered four different deployment scenarios for launching the OTA impersonation attack: (i)   \textit{Wired Attack Scenario}, where the attacker hardware sends the impersonated signals over a wired link (see Figure~\ref{fig:Testbed}(c)), yielding an SNR of approximately $38$dB. (ii) \textit{LoS Wireless Attack Scenario}, where the attacker transmits from a LoS distance of $1.5$m from RX$_V$ (see Figure~\ref{fig:Testbed}(d)), resulting in an SNR of about $26$dB. (iii) 
\textit{NLoS Wireless Attack Scenario}, where, while keeping $1.5$m separation between RX$_A$ and RX$_V$, a large metal reflector is placed in the middle, forcing the signal to reach the receiver primarily through multipath reflections. This configuration yields an SNR in the range of $13.5-16$dB. (iv) \textit{LLoS Wireless Attack Scenario}, where the attacker launches its attack $7$m away from RX$_V$ in a rich multipath environment with rich multipath components (see Figure~\ref{fig:Testbed}(e)). This scenario results in an SNR range of $9-13$dB.

\section{Direct Injection Attack Results}
\label{sec:results}
This section presents a comprehensive evaluation of the proposed attack, when the crafted signals are directly fed to the victim RFFP model.
For this, we used 7 different random seeds to train 7 victim RFFP classifiers using the data captured over wired transmission via RX$_V$-Ch1. In parallel, we train another 7 independent surrogate classifiers on the attacker's hardware for each corresponding dataset captured via RX$_A$, following the \textit{No-Access} threat model. This results in 49 testing instances when cross-testing these models, which we present their average and standard deviation. 
The RFFP classifiers employ the \textbf{CNN1} architecture with the \textbf{RawIQ} input representation, as detailed in Appendix~\ref{apn:CLSs}. 
More advanced experiments, involving mismatched DL architectures and more robust feature representations, will be presented in the following sections.
The metric we used for our evaluation is the Targeted Attack Success Rate ({TASR}), defined as the percentage of our crafted signals impersonating a target device that are identified by the victim RFFP classifier as that target device. 

\subsection{Robustness and Scalability} \label{subsec:Robustness}
We begin by comparing the proposed frameworks against the following existing targeted impersonation attack baselines:  

\begin{enumerate} [noitemsep, topsep=0pt, leftmargin=*]
\item \textbf{Projected Gradient Descent (PGD)~\cite{kurakin2018adversarial}:}  PGD iteratively adds adversarial perturbations to clean signals, following Eq~\eqref{eq:PGD}, without first embedding the estimated impairments. We experiment with different $\epsilon$ values, while $\alpha=\epsilon/50$.


\item \textbf{Collusion-Driven Impersonation Attack (CDIA)~\cite{xu2025collusion}:}
This approach was recently proposed for impersonation attacks, where a variational autoencoder (VAE)-based generative network is trained to replicate the target device's RF fingerprint. Our implementation adheres to the original architecture and training parameters described in that work.

\item \textbf{Random Signal:}
We also include a random-guess baseline to better contextualize the task's difficulty. Since our setting assumes a closed-set classifier that must assign each input to one of the $K$ known devices, the theoretical attack success rate under random input is $\text{TASR} = 1/K$.

\end{enumerate}




\begin{figure}
    \centering
    \includegraphics[width=0.9\linewidth]
    {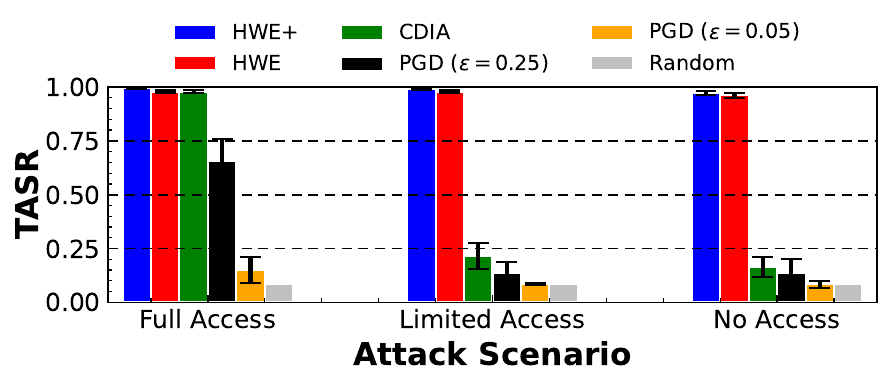}
    \caption{TASR for different baseline methods.}
    \label{fig:DI_12}
\end{figure}

Figure~\ref{fig:DI_12} presents the TASR impersonating 12 devices obtained under each of the three access models: \textit{Full-Access}, \textit{Limited-Access} and \textit{No-Access}. 
The figure shows that applying \textbf{PGD} with a small distortion of $\epsilon=0.05$ fails completely under all access models, and does only a few percentages better than random guessing. On the other hand, increasing the distortion to $\epsilon=0.25$ raises TASR to around 63\%, but only under the unrealistic \textit{Full-Access} model.
It is also worth highlighting that at $\epsilon=0.25$, BER$=14.63\%$, as already illustrated in Figure~\ref{fig:Motivation}, making this approach impractical for RFFP impersonation attacks, since the communication decoding capability will be compromised.
The \textbf{CDIA} approach shows excellent performance under the \textit{Full-Access} model, but fails significantly under the other two (more realistic) models. Finally, our approaches (\textbf{HWE} and \textbf{HWE+}) both outperform the baselines significantly, achieving a TASR of more than $97\%$ consistently across all three access models.

\begin{figure}
    \centering
    \includegraphics[width=0.9\linewidth]{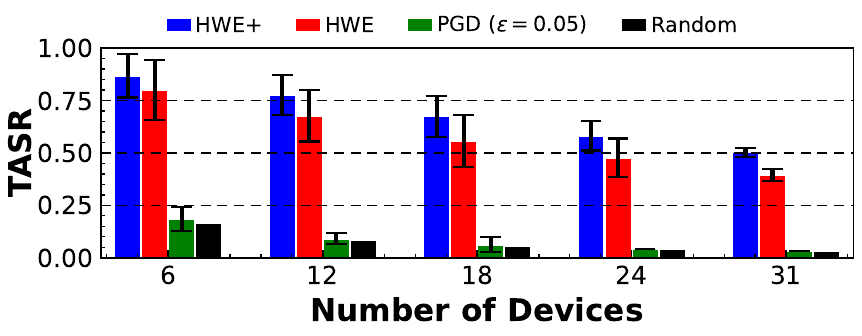}
    \caption{Scalability results under the \textit{No-Access} model.}
    \label{fig:DI_Scalability}
\end{figure}

After proving the effectiveness of the proposed attacks across all access models,
we evaluate their scalability and generalization by varying the number of enrolled devices, specifically, when setting $K = 6, 12, 18, 24$, or $31$. 
%
For each $K$, we randomly sample 30 distinct device subsets of size $K$, train a separate model for each subset, and report the average TASR across all runs. All experiments here are conducted under the \textit{No-Access} model. 
As shown in Figure~\ref{fig:DI_Scalability}, TASR exhibits a expected monotonic decline as $K$ increases.
This trend is consistent with prior findings that legitimate RFFP classifiers begin to experience increased confusion among classes when the number of enrolled devices grows~\cite{albousayri2025bluetooth}. 
This is because the likelihood of having devices with overlapping levels of distortion--such as Dev4, Dev5, and Dev10 in Figure~\ref{fig:Impairments-vis}--increases.
Overall, the proposed approaches maintain high TASR across all scales, with minimum values of $39.6\%$-$50.3\%$ at $K=31$. 
These results imply that the attacker could easily compromise such systems under a relaxed, untargeted attack setting, where the goal is to impersonate \textit{any} valid device in the set rather than a specific one.

\subsection{Robustness to Input Data Representations}\label{subsec:results_techs}

Here, we evaluate the effectiveness of our attack framework when the target (victim) RFFP system employs alternative feature representations rather than normalized raw IQ frames as classifier inputs. Specifically, we consider the following feature-representation baselines:

\begin{enumerate} [noitemsep, topsep=0pt, leftmargin=*]
\item \textbf{Phase Derivative (PD)~\cite{albousayri2025bluetooth}:}
PD uses the first-order derivative of the IQ phase as an input to the classifier, defined as $\text{PD}[n]=\Delta_n\angle_ux[n]$. 
PD achieved robust performance in RFFP domain generalization by suppressing channel noisy phase while preserving key impairments.

\item \textbf{Difference of the Logarithm of the Spectrum (DoLoS)~\cite{xing2022design}:}
DoLoS assumes channel coherency within $x[n]$ and leverages it to mitigate the channel impact. It divides the signal into two non-overlapping sequences $x^{L}[n] = x[n]$  and $x^{R}[n] = x[n + N/2]$ for $ 0 \leq n < N/2$. 
The feature representation is then defined as:
$\text{DoLoS}[n] = \log_{10}(|\text{FFT}({x^{R}[n]})|) - \log_{10}(|\text{FFT}({x^{L}[n]})|)$.

\item \textbf{Centralized Logarithmic Power Spectrum (CLPS)~\cite{tang2024causal}:} 
CLPS removes global spectral bias introduced by the channel, and defined as
$\text{CLPS}[n] \!=\! \log_{10}(|\text{FFT}(x[n])|^2)-\langle \log_{10}(|\text{FFT}(x[n])|^2)\rangle$, with $\langle \cdot \rangle$ being the mean.


\item \textbf{Magnitude, Phase and FFT Feature Extraction (Mbed)~\cite{jagannath2023embedding}}:
Mbed has been shown to outperform raw IQ based classification by concatenating both the time and frequency domain features. It is defined as:
$\text{Mbed}[n] = [ |{x}[n]|,$ $ \angle {x}[n]$, $ |\text{FFT}(x[n])|^2]^T$.
\end{enumerate}

\begin{figure}
    \centering
    \includegraphics[width=0.9\linewidth]{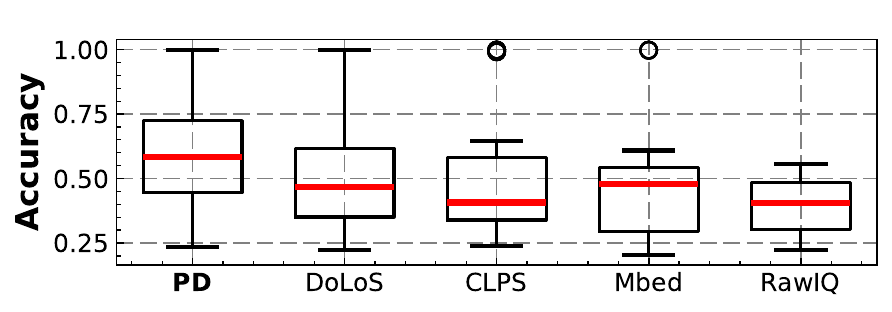}
    \caption{Classification accuracy for different baselines.}
    \label{fig:31CLS_2}
\end{figure}

Before presenting the attack results under the studied baseline input data representations, we first examine how these baseline representation techniques perform in device classification accuracy.
Figure~\ref{fig:31CLS_2} plots the average classification accuracy of these baselines, when deployed under different scenarios (e.g., different locations and/or channels).
The figure indicates that \textbf{PD} yields the best performance, followed by \textbf{DoLoS} then \textbf{CLPS} and finally \textbf{Mbed} and \textbf{RawIQ}.

%

\begin{figure}
    \centering
    \includegraphics[width=0.9\linewidth]{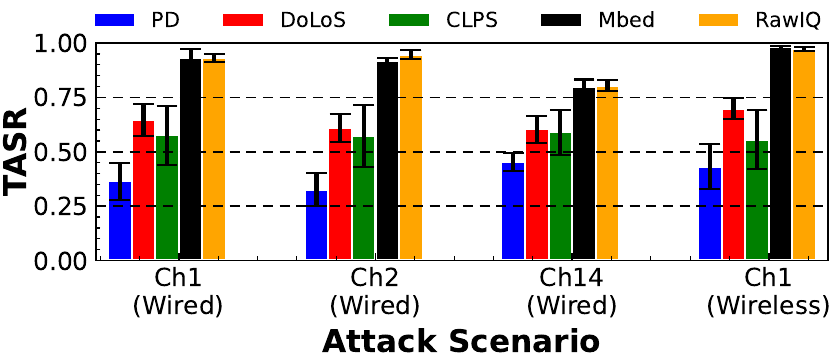}
    \caption{Performance under different data representations and across different domains (channel and wired/wireless).}
    \label{fig:TASR_AcrossTECHs}
\end{figure}

In our assessment of the robustness of the proposed attack framework against different input data representations, we first train the victim classifier using the studied representation technique and then launch the impersonation attack. In all scenarios, the data used by the victim receiver to train its classifier is collected using Ch1 over a wired setup.
At the attacker, impersonation signals are generated via the proposed attack framework, \textbf{HWE+}, under the \textit{No-Access} assumption using data collected from Ch1, Ch2, and Ch14 under the wired setup and Ch1 under the wireless setup. These signals are then injected into the victim classifier.
The TASR results are shown in 
Figure~\ref{fig:TASR_AcrossTECHs} when impersonating 12 devices, which 
clearly show that using \textbf{RawIQ} as input features---matching the attack classifier's input---yields the highest TASR. In contrast, the lowest TASR is observed when the legitimate classifier employs \textbf{PD} features. Both \textbf{DoLoS} and \textbf{CLPS} representations yield comparable mid-level performance, whereas \textbf{Mbed} consistently achieves high TASR across all setups.
From a defense perspective, these results indicate that using \textbf{PD} features provides strong robustness against this class of impersonation attacks, while simultaneously maintaining high classification accuracy at the legitimate RFFP system.

\subsection{Generalizability to Unseen Training Sequences} 
\begin{figure}
    \centering
    \includegraphics[width=0.9\linewidth]{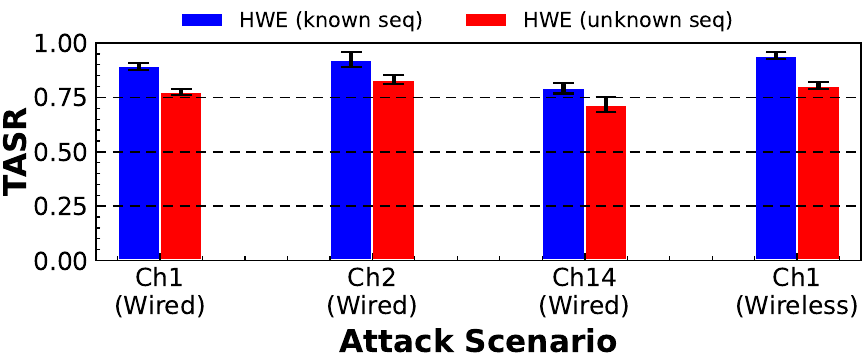}
    \caption{Robustness to unknown inference sequence.}
    \label{fig:TASR_Unseen_bits}
\end{figure}

Unlike prior methods, we argue that our \textbf{HWE} approach extracts accurate estimates of impairments using the preamble portion. When used to generate the adversarial frame, these estimates preserve a consistent RFFP identity throughout the entire frame.
This is particularly important because legitimate/victim RFFP systems may authenticate devices using hidden or internal sequences within the BLE frame, rather than a fixed, known segment. 
Here, we evaluate the robustness and consistency of our attack against victim classifiers that use sequence that is unknown (hidden) to the attacker.  
To do so, we divide the captured signal $x_T[n]$ of length $N=220$ into two halves: the preamble $x_T^{L}[n]$ and the remaining portion $x_T^{R}[n]$. Specifically, the attacker uses only $x_T^{L}[n]$ to estimate the impairment vector and embeds it to generate the synthesized signal $x_A[n]$, that is impersonating $x_T[n]$.
In contrast, the victim classifier is trained to perform identification solely based on $x_T^{R}[n]$, a segment never used by the attacker. 
Figure~\ref{fig:TASR_Unseen_bits} compares the TASR when the victim relies on an inference sequence unknown to the attacker (i.e., $x_T^{R}[n]$) against the baseline case where both the attacker and the victim use the same sequence (i.e., $x_T^{L}[n]$).
The results demonstrate consistently high TASR, even when the portion used for training and inference is unknown to the attacker---denoted as {\textbf{HWE (unknown seq)}} in the figure. In this case, the performance degradation remains modest, with an average drop ranging from $8.5\%$ to $13.7\%$.

\section{Over-the-Air (OTA) Attack Results}
\label{sec:results2}
\begin{figure}[t]
    \centering
    \includegraphics[width=0.9\linewidth]{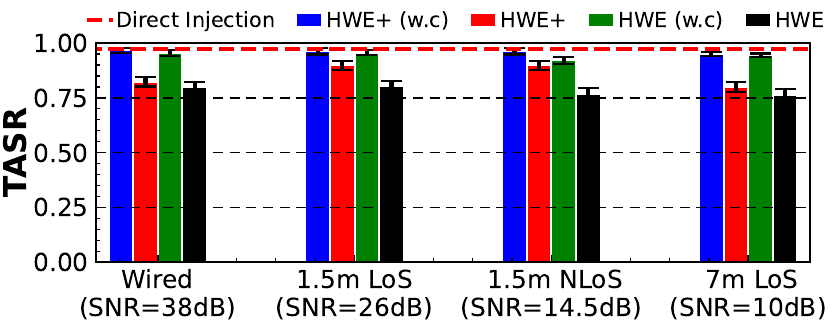}
    \caption{OTA attack results. w.c. refers to 'with calibration'.}
    \label{fig:TASR_OTA}
\end{figure}
After demonstrating the effectiveness of our framework under direct injection setup, we now evaluate it under the OTA settings, where crafted attack signals are modulated and transmitted to the victim receiver RX$_V$ using the attacker's hardware RX$_A$. The OTA evaluation includes wired and wireless links under LoS and NLoS conditions, as described in Section~\ref{subsec:OTAsetup}.
\subsection{Impact of OTA Distortions}
When the impersonation attack is launched using the OTA strategy, the impersonated OTA signals go through the attacker's transmitter hardware RX$_A$, propagation channel, and then the victim's receiver hardware RX$_V$, resulting in inevitable signal distortions. The severity of these distortions is highly correlated with the quality of the hardware, where low-end transceivers usually yield more distortions~\cite{rehman2014analysis}.
Our measurements reveal that these distortions are around $870$Hz for $f_{CFO}$ and around $1.2\!\times\!10^{-5}$ and $-1.6\!\times\!10^{-6}$ for $I_{DC}$ and $Q_{DC}$, respectively, with negligible values for the rest of the impairments.  
Once those OTA-incurred distortions are measured, they can then be calibrated in the digital baseband domain to mitigate their impact. For evaluation, we implemented the OTA attacks both with and without calibration of the measured additional distortions.
We show in Figure~\ref{fig:TASR_OTA} the TASR values achieved under the OTA attack strategy, with and without calibration, while varying the distance between the attacker and the victim from $1.5$m to $7$m. 
The figure shows that \textbf{HWE+} outperforms \textbf{HWE} in all cases, with a monotonic decrease in TASR as the SNR decreases. Overall, calibrating the induced impairments increases TASR by up to $20\%$, achieving performance that is closely aligned with the Direct Injection strategy, as can be seen from Figure~\ref{fig:TASR_OTA}.

\begin{figure}
    \centering
    \includegraphics[width=0.9\linewidth]{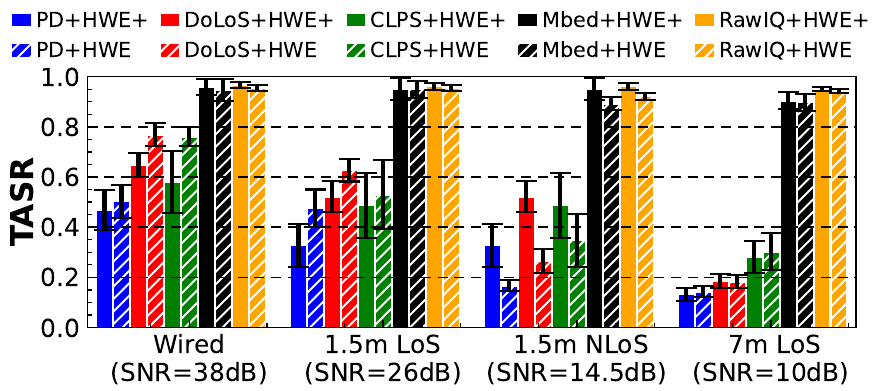}
    \caption{Robustness to different input representations.}
    \label{fig:TASR_OTA2}
\end{figure}
\subsection{Robustness to Input Data Representations}
In this experiment, we evaluate the TASR when the captured OTA signals are fed into victim RFFP classifiers that use different input data representation baselines. The same data representation baselines described and used in Section~\ref{subsec:results_techs} are used in this section, and the results for \textbf{HWE+} and \textbf{HWE} attacks with calibration of excess impairments are reported.
Figure~\ref{fig:TASR_OTA2} shows that both \textbf{RawIQ} and \textbf{Mbed} representations are relatively easier to attack across all setups. In contrast, \textbf{PD}, \textbf{DoLoS}, and \textbf{CLPS} exhibit a monotonic decrease in TASR as the SNR increases, with \textbf{PD} consistently achieving the lowest TASR. The figure further indicates that under wired and LoS conditions, \textbf{HWE} slightly outperforms \textbf{HWE+}, whereas the opposite trend is observed under NLoS conditions.

\subsection{Robustness to Deep Learning Architectures}
\begin{figure}
    \centering
    \includegraphics[width=0.9\linewidth]{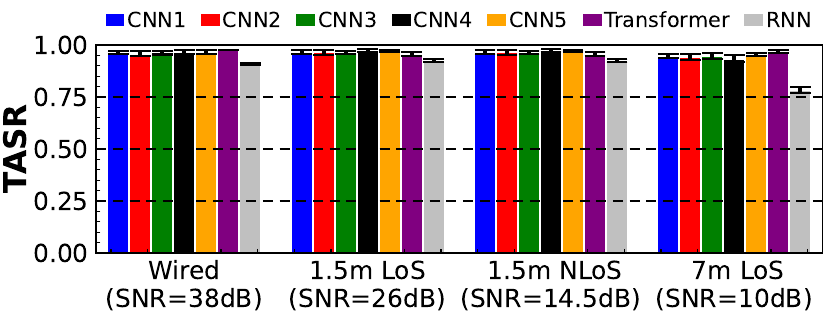}
    \caption{Robustness to victim's DL architecture models.}
    \label{fig:TASR_OTA3}
\end{figure}
This section study the impact of having mismatched DL-RFFP architecture between the victim system and the surrogate classifier trained at the attacker side. This scenario is particularly important in practice, as the architecture of the victim system is typically unknown, making matched designs unlikely. Since \textbf{HWE+} explicitly relies on the surrogate classifier to refine and craft the attack signal, only its results are relevant and therefore reported.
In this experiment, the attacker's classifier was kept fixed using \textbf{CNN1} configuration, while varying the architecture of the legitimate/victim classifier by considering and evaluating: five different CNN architectures, one Transformer-based model, and one Recurrent Neural Network (RNN)-based classifier. Details about the configuration of each of these studied architectures can be found in Appendix~\ref{apn:CLSs}.

Figure~\ref{fig:TASR_OTA3} shows that our framework remains robust across different victim model architectures, achieving nearly the same TASR as in the Direct Injection strategy.
An interesting observation is that matched DL architecture models does not always imply best performance, as this can be seen when the attack signals are evaluated over \textbf{CNN4},  \textbf{CNN5} and \textbf{Transformer} architecture, achieving around $0.72\%$ increase over \textbf{CNN1}. 
In contrast, evaluating the attack over \textbf{RNN} results in the lowest TASR and appears to be more sensitive to low SNRs.

\section{Conclusion}
\label{sec:conclusion}
We present a practical impersonation framework against DL-based RFFP systems. Unlike prior approaches that rely on unrealistic assumptions, our method explicitly synthesizes device-specific hardware impairments. By accurately estimating these impairments, we generate physically consistent spoofed signals matching the target device's intrinsic RF characteristics. Evaluations demonstrate our attack's robustness and transferability across unseen setups, revealing a fundamental vulnerability in existing DL-based RFFP approaches.


\bibliographystyle{IEEEtran}
\bibliography{sample-base}

\appendices

\section{Setup and Training Parameters}
The dataset is split per device into 2000 training, 400 testing, and 200 validation frames for classification, leading to 62,000 unlabeled training, 12,400 testing, and 6,200 validation frames for $\mathbf{HWE}(\cdot;\Phi)$ block. We use the first \(N=220\) samples of each frame under the \textbf{RawIQ} representation, forming power-normalized IQ windows of shape \((2,N)\). All models are trained for 50 epochs using AdamW with a batch size of 128.

\label{apn:HWEblock}
The $\mathbf{HWE}$ block (67.6K parameters) is implemented using a CNN with four 1D convolutional layers (filter sizes \(F=\{16,64,96,8\}\), kernel sizes \(H=\{9,11,3,20\}\)), each followed by SiLU activation and max-pooling (size 2). The output is flattened and passed through two fully connected (FC) layers of sizes \(\{128,64\}\) with Leaky ReLU activations, followed by an output layer with \(|\Theta|\) neurons to predict \(\hat{\Theta}\).
Training uses Mean Absolute Error (MAE), which outperformed MSE, with phase regularization parameters \(\lambda^{\text{max}}_{\text{Phase}}=10\) and \(\lambda^{\text{min}}_{\text{Phase}}=0.001\).
\label{apn:CLSs}
We consider several DL classifiers, including five CNN variants and RNN- and Transformer-based models. The CNNs share a common 1D convolutional backbone with max-pooling and FC layers, differing in depth and width: \textbf{CNN1} (1.34M) is the baseline with filter sizes \(F=\{32, 64, 96, 96, 96\}\) and kernel sizes \(H=\{24,12,6,3,3\}\), followed by 2 FC layers of \(\{128,64\}\) neurons; \textbf{CNN2} (1.69M) removes the last convolutional block and uses a smaller final FC layer; \textbf{CNN3} (0.89M) further reduces FC size; \textbf{CNN4} (0.71M) adds an extra convolutional block with smaller FC layers; and \textbf{CNN5} (10.02M) is a larger variant with deeper filters and wider FC layers. We also include a \textbf{Transformer} model (3.69M) with 32-dimensional embeddings, a 2-layer encoder with 4 attention heads and MLP size 512, and a 512-neuron FC layer, as well as a \textbf{GRU-based RNN} (0.83M) with three bidirectional layers (hidden size 128), followed by LayerNorm and a 512-neuron FC layer. All models use an output layer with \(K\) neurons (number of devices).

\end{document}